\documentclass[11pt,a4paper]{article}

\usepackage{jheppub}

\usepackage{amsmath}
\usepackage{amssymb}
\usepackage{graphicx}
\usepackage{float}
\usepackage{braket}
\usepackage{hyperref}

\usepackage[all]{hypcap}

\usepackage{feynmp}
\newcommand{\bo}{\mathrm{O}}
\newcommand{\op}{\mathcal{O}}
\newcommand{\Tev}{\, \mathrm{TeV}}
\newcommand{\Gev}{\, \mathrm{GeV}}
\newcommand{\ewgg}{$SU(2) \times U(1)$ }
\newcommand{\ewggnospace}{$SU(2) \times U(1)$}
\newcommand{\msbar}{$\overline{MS}$}

\newcommand{\dif}{\mathrm{d}}

\newcommand{\dfk}{\frac{\dif^4 k}{(2\pi)^4} \,}
\newcommand{\dfkp}{\frac{\dif^4 k^\prime}{(2\pi)^4} \,}
\newcommand{\pt}{\partial}
\newcommand{\al}{\alpha}
\newcommand{\del}{\delta}
\newcommand{\lag}{\mathcal{L}}
\newcommand{\Tr}{\mathrm{Tr}}
\newcommand{\abs}[1]{\lvert #1 \rvert}
\newcommand{\La}{\Lambda}

\usepackage{color}

\newcommand{\kz}{\alpha_Z}
\newcommand{\kg}{\alpha_\gamma}
\newcommand{\kb}{\alpha_B}
\newcommand{\kw}{\alpha_W}

\title{On LHC searches for $CP$-violating, dimension-6 electroweak gauge boson operators}
\author[a]{Ben Gripaios}
\author[a]{Dave Sutherland}

\affiliation[a]{Cavendish Laboratory, \\ J.J.\ Thomson Avenue, Cambridge, UK}

\emailAdd{gripaios@hep.phy.cam.ac.uk}
\emailAdd{dws28@cam.ac.uk}

\preprint{Cavendish-HEP-13/08}

\abstract{We reconsider the prospects for observing a dimension-6, $CP$-violating operator involving $W^+W^-Z$ at the LHC. 
Firstly, we correct a number of earlier calculations 
of the loop contribution to the 
the neutron electric dipole moment of a companion operator, involving $W^+W^-\gamma$, showing that measurements imply a very strong bound on the companion operator.
Secondly, we quantify the link between the two operators, showing that strongly-coupled new physics could only be observable in proposed searches if
it appeared at a scale below $\sim 170$ GeV. This is most unlikely, given the null results of other searches at the LHC and elsewhere.}

\begin{document}
\maketitle

\section{Introduction}
Despite its laudable performance, the first run of the Large Hadron Collider (LHC) saw no evidence for physics beyond the Standard Model, putting the naturalness paradigm under severe pressure. This has the twofold effect of pushing the bounds on the new physics scale higher and making 
theorists' rhetoric about what we should be looking for less convincing. In light of this, it makes increasing sense for experiments to frame their searches
in terms of effective Lagrangians, in which new physics is parameterised by higher dimension operators built out of the Standard Model degrees of freedom. Even if no new physics is found (alas!), this approach will ensure that the LHC leaves a useful legacy in its wake, in the form of optimal, model-independent constraints on possible new physics.

Of particular interest (independently of the naturalness issue) are higher dimensional operators violating $CP$, which could generate the baryon asymmetry in the Universe. In this work, we examine one such operator,\footnote{As far as we are aware, no-one has yet suggested that this operator plays a r\^{o}le in baryogenesis.} namely $\op_Z \equiv {W^+}^\mu_\nu {W^-}^\nu_\lambda \tilde{Z}^{\lambda}_\mu$ where ${W^\pm}^\mu_\nu$ is the usual field strength tensor for $W^\pm$ and $\tilde{Z}_{\mu\nu} \equiv \epsilon_{\mu\nu\rho\sigma} Z^{\rho\sigma}$ is the dual field strength tensor for the $Z$. $\op_Z$ has been suggested more than once as a suitable target for LHC searches \cite{han,wells}.\footnote{Refs. \cite{han} and \cite{wells} differ in that the former proposes a search based on observables that are genuinely odd under $CP$, while the latter's observables are odd only under the ``na\"{i}ve'' time-reversal transformation (which reverses momenta and spins \cite{han}), meaning that effects could be generated by $CP$-conserving physics in the presence of, {\em e.g.}, final state interactions.}
We shall see that, for a variety of reasons, there is almost no hope of being sensitive to allowed values of the coefficient, $\kz$, of this operator at the LHC, and so experimentalists would be better off directing their efforts elsewhere. Along the way, we shall have occasion to point out one or two pitfalls in the logic of effective field theory (EFT) that, while no doubt known to many, are liable to trap the unwary. 

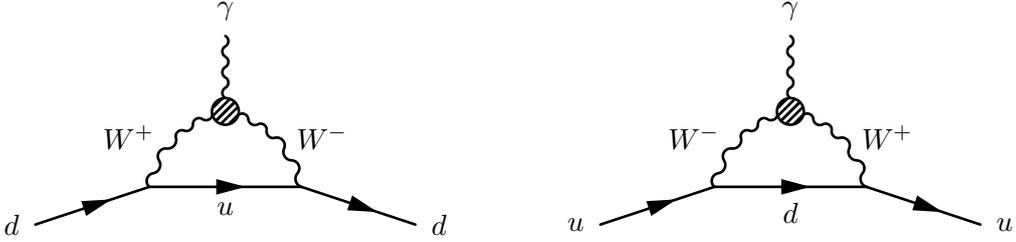
\begin{figure}
\centering
\begin{tabular}{c c}

\begin{fmffile}{downedm_feynmp}
\begin{fmfgraph*}(50,50)
\fmfleft{din}
\fmfright{dout}
\fmftop{gamma}
\fmf{fermion}{din,v1}
\fmf{fermion,label=$u$,tension=0.5}{v1,v2}
\fmf{fermion}{v2,dout}
\fmf{photon,left=0.3,tension=0.5}{v1,vblob,v2}
\fmf{photon}{gamma,vblob}
\fmfblob{10}{vblob}
\fmflabel{$d$}{din}
\fmflabel{$d$}{dout}
\fmflabel{$\gamma$}{gamma}
\fmffreeze
\fmfiv{l=$W^-$,l.a=-20,l.d=.2w}{vloc(__vblob)}
\fmfiv{l=$W^+$,l.a=200,l.d=.2w}{vloc(__vblob)}
\end{fmfgraph*}
\end{fmffile}
\hspace{15mm}
&
\begin{fmffile}{upedm_feynmp}
\begin{fmfgraph*}(50,50)
\fmfleft{din}
\fmfright{dout}
\fmftop{gamma}
\fmf{fermion}{din,v1}
\fmf{fermion,label=$d$,tension=0.5}{v1,v2}
\fmf{fermion}{v2,dout}
\fmf{photon,left=0.3,tension=0.5}{v1,vblob,v2}
\fmf{photon}{gamma,vblob}
\fmfblob{10}{vblob}
\fmflabel{$u$}{din}
\fmflabel{$u$}{dout}
\fmflabel{$\gamma$}{gamma}
\fmffreeze
\fmfiv{l=$W^+$,l.a=-20,l.d=.2w}{vloc(__vblob)}
\fmfiv{l=$W^-$,l.a=200,l.d=.2w}{vloc(__vblob)}
\end{fmfgraph*}
\end{fmffile}

\end{tabular}

\caption{One-loop contributions of $\op_\gamma = {W^+}^\mu_\nu {W^-}^\nu_\lambda \tilde{F}^{\lambda}_\mu$ (shaded blob) to the neutron EDM. Custodial symmetry implies that the diagrams change sign under $m_u \leftrightarrow m_d$. \label{fig:nedm}}
\end{figure}

In a nutshell, the reason why there is no hope of being sensitive to $\kz$ at the LHC is that $\op_Z$ is linked in the Standard Model (in a way that we shall make precise) to the operator $\op_\gamma \equiv {W^+}^\mu_\nu {W^-}^\nu_\lambda \tilde{F}^{\lambda}_\mu$, which contributes to the electric dipole moment (EDM) of the neutron via the 1-loop diagrams in Fig.~\ref{fig:nedm}.\footnote{The operator $\op_Z$ also contributes directly to the EDM via 2-loop diagrams, but we do not consider them here.} Two questions thus arise: Firstly, how large is the contribution of $\op_\gamma$ to the EDM? Secondly, how precise is the link between $\op_\gamma$ and $\op_Z$?

In answer to the first question, we have found five independent computations in the literature (see Table~\ref{tab:nedmcalcs}) of the diagrams of Fig.~\ref{fig:nedm}; no two sets of authors agree on the result, with one set \cite{novalessanchez} suggesting a suppression by a factor $\sim 10^{-10}$, compared to a na\"{\i}ve estimate. Another group of authors \cite{boudjema} could even be said to not agree with themselves, in that their result depends on how the regularization is performed, and in fact can take any value. They conclude, therefore, that the result must be intrinsically dependent on the specific nature of the UV physics that generates the EFT operator. Such a result, if correct, would sound the death knell for quantum field theory, which Wilson showed us was nothing more than a consistent paradigm for organizing the quantum contributions of physics on differing time scales. Specifically, physics at a given energy scale retains all information about physics at higher scales, but that information is carried only in the values of coupling constants multiplying EFT operators, which can be organized in an expansion in energies.
The implication of \cite{boudjema} is that information is not passed down in this way: to wit, the physics of the neutron EDM somehow requires information about the UV that is not encoded in the effective theory at the weak scale. In Section~\ref{sec:nedmcalc} we explain all these disagreements and provide a sixth computation of our own, which has the merit of agreeing with one of the existing computations, namely \cite{boudjema}, when the latter is interpreted correctly. There is no suppression of the EDM.

As for the second question, the strength of the link between $\op_Z$ and $\op_\gamma$ appears to depend on whether or not we insist that the higher dimension operators of the theory respect the full $SU(2)\times U(1)$ invariance of the SM. If we do, then both operators descend from the $SU(2)\times U(1)$ invariant, dimension-six operator $\op_W \equiv {W^+}^\mu_\nu {W^-}^\nu_\lambda \tilde{W}^3{}^\lambda_\mu$ and the coefficients are related by $c_W \kg = s_W \kz$.\footnote{To make the gauge invariance more manifest, write the operator as $\op_W = {W^+}^\mu_\nu {W^-}^\nu_\lambda \tilde{W}^3{}^\lambda_\mu = i {W^1}^\mu_\nu {W^2}^\nu_\lambda \tilde{W}^3{}^\lambda_\mu \subset \frac{i \epsilon^{abc}}{3!} {W^a}^\mu_\nu {W^b}^\nu_\lambda \tilde{W}^c{}^\lambda_\mu$.} The strong constraint from the neutron EDM then implies that there is no sensitivity to $\kz$ at the LHC. But it has long been customary for authors to tacitly adopt the view (see, {\em e.g.}, \cite{Hagiwara:1986vm} and myriad references thereto) that, since $SU(2)\times U(1)$ is ultimately broken, there is no need to insist that this symmetry be respected by higher dimension operators in the theory. If so, one can freely add either of the ($U(1)_{\mathrm{em}}$-preserving) operators $\op_W$ or $\op_B = {W^+}^\mu_\nu {W^-}^\nu_\lambda \tilde{B}^{\lambda}_\mu$ to the Lagrangian;\footnote{That this is not $SU(2)_L$ invariant is best seen by writing $\op_B = {W^+}^\mu_\nu {W^-}^\nu_\lambda \tilde{B}^\lambda_\mu = i {W^1}^\mu_\nu {W^2}^\nu_\lambda \tilde{B}^\lambda_\mu$, whereas the only two-field-strength invariant is $W^a W^b \del^{ab} = W^a W^a$. An operator involving $W^a W^a$ vanishes identically: ${W^a}^\mu_\nu {W^a}^\nu_\lambda \tilde{B}^\lambda_\mu = (-{W^a}^\nu_\mu) (-{W^a}^\lambda_\nu) (-\tilde{B}^\mu_\lambda) = -{W^a}^\lambda_\nu {W^a}^\nu_\mu \tilde{B}^\mu_\lambda = 0$.} a non-vanishing coefficient, $\kb \neq 0$, of $\op_B$ violates the relation $c_W \kg = s_W \kz$ and so a bound on $\kg$ from the neutron EDM no longer implies a bound on $\kz$. 

Most readers, we hope, will feel at least a vague aesthetic unease at this custom. After all, if we do not need to respect $SU(2)\times U(1)$ in higher dimension operators, why do we bother respecting it in the lower dimension operators of the Standard Model Lagrangian? As physicists, our duty is to to replace this aesthetic unease with quantitative objections. 
A robust attack in this direction is begun in \cite{derujula}, with the {\em pronunciamento} ``The standard model {\em is} [{\em sic}], and new hypothetical dynamics {\em must be}, $SU(2)\times U(1)$ gauge-invariant above the symmetry breaking scale. Else, it is difficult to imagine how the theory would maintain its sacred renormalizability or how symmetry breaking tell-tale signatures would not percolate down to observable levels, particularly in those observables that are understood up to (finite, calculable) electroweak radiative corrections.'' But it is not clear to us what r\^{o}le renormalizability plays,
given that the EFT does not possess it (nor does any UV completion, necessarily). Furthermore, at least for the operators $\op_W$ and $\op_B$ considered here, we shall see that the corrections to precision electroweak observables turn out to be small.

In Section~\ref{sec:unitcutoff}, we exhibit a quantitative problem of a different kind. As is well known, an EFT is valid only in some regime of energy, up to a cut-off scale $\Lambda$, beyond which either its perturbative expansion ({\em ergo} its predictive power) or consistency breaks down. In our view, questions of aesthetics are then best rephrased as the question: what is $\Lambda$? 
An aesthetically beautiful EFT (strictly speaking, a formulation thereof, since the same physics can be described by many Lagrangians), is then one in which the value of $\Lambda$ not only is non-zero, but also can be seen directly by the beholder of the Lagrangian. In contrast, an EFT where 
$\Lambda$ is not manifest, and turns out to be lower than one might na\"{\i}vely guess, may be considered less attractive. At its most beastly, an EFT may turn out to have $\Lambda = 0$, in which case the alternative moniker of `{\em in}effective field theory' would seem more apposite.

The EFT containing $\kb \op_B$, with $\kb \equiv \La_B^{-2}$ falls in the category of `less attractive' (to use the the vernacular):
the cut-off is not $\La_B$, but rather scales as $\sqrt{v \La_B}$, where $v$ is the electroweak scale. This is obvious in a formulation that appears more beautiful, but which is in fact completely equivalent, in which $SU(2)\times U(1)$ is manifest, with deviations from the relation $c_W \kg = s_W \kz$ arising from operators beginning at dimension eight. 
But it can also be seen easily enough in the original formulation in terms of operators $\op_\gamma$ and $\op_Z$.
Moreover, it is evident (in either formulation) that the cut-off is lowered by a further factor. As a result, we conclude that visible effects of the operator $\op_Z$ at the LHC require a new physics scale around $170 \Gev$. If there did exist strongly coupled physics at such a low scale, effects would appear all over the place at the LHC. Dedicated searches of the type advocated would be superfluous.

\section{The 1-loop contribution of $\op_W$ to the neutron EDM}
\label{sec:nedmcalc}
We begin with the question of whether the \ewgg covariant operator $\op_W = {W^+}^\mu_\nu {W^-}^\nu_\lambda \tilde{W}^3{}^\lambda_\mu$ could be detectable at the LHC. The operator is odd under $CP$, but invariant under the group of flavour symmetries of the SM and so an immediate concern is whether it can avoid bounds on flavour-singlet $CP$-violation from EDM measurements, notably that of the neutron.\footnote{$CP$-odd operators such as $\op_W$ and $\op_B$ also
contribute to $CP$-even electroweak precision observables, via diagrams containing $\geq2$ insertions.
For example, $\op_B$ gives contributions to $g^2 \hat{U}$ and $g^{-2} m_W^2 \hat{T}$ (defined as in \cite{barbieri}) of order
$\kb^2 \frac{m_W^4}{16 \pi^2}$ and $\kb^2 g^2 \frac{m_W^6}{(16 \pi^2)^2}$, respectively, implying weak bounds of $\kb \lesssim (100 \Gev)^{-2}$ and $\kb \lesssim (20 \Gev)^{-2}$, respectively. Note that the bound from $\hat{T}$ is unusually poorly constraining as $\op_B$'s contribution must be 2-loop (any 1-loop diagram of two $\op_B$s has derivatives on the external legs). $\op_W$'s contribution to $2 g^{-2} m_W^{-2} W$ of $\sim \kw^2 \frac{m_W^2}{16 \pi^2}$ gives a similarly loose limit of $\kw \lesssim (90 \Gev)^{-2}$.}

\begin{table}
\centering
\begin{tabular}{c | c | c}
Authors & Regularization & $d_f$ \\
\hline
Atwood \emph{et al.} \cite{atwood} & cut-off $\Lambda$ & $ m_f \kg \frac{g^2}{64 \pi^2} [ \ln \left( \frac{\Lambda^2}{m_W^2} \right) + \mathrm{O}(1) ] $ \\
Boudjema \emph{et al.} \cite{boudjema} & $\overline{MS}$ & $ m_f \kg \frac{g^2}{64 \pi^2}$ \\
Hoogeveen \cite{hoogeveen} & cut-off $\Lambda$& $0$ \\
Novales-S\'anchez \& Toscano \cite{novalessanchez} & $\overline{MS}$ & $ m_f \kg \frac{g^2 s_W}{64 \pi^2} \cdot \frac{2m_f^2}{3m_W^2}$ \\
de R\'ujula \emph{et al.} \cite{derujula} & cut-off $\Lambda$& $ m_f \kg \frac{g^2}{64 \pi^2} \frac{2}{s_W^2} \ln \left( \frac{\Lambda^2}{m_W^2} \right) $ \\
\end{tabular}
\caption{The effective operator $ - \frac{1}{2} d_f \overline{\psi} \sigma^{\mu \nu} \psi \tilde{F}_{\mu \nu} $ for a down quark $\psi$ of mass $m_f$ due to a 1-loop diagram including the operator $ - i \kg {W^+}^\mu_\nu {W^-}^\nu_\lambda \tilde{F}^\lambda_\mu $. The sign of the result is reversed for an up quark.}
\label{tab:nedmcalcs}
\end{table}

As described in the Introduction, none of the five groups attempting this calculation are in agreement on the result. The relevant 1-loop diagrams are depicted in Fig.~\ref{fig:nedm}, and we list the results of the five calculations in Table~\ref{tab:nedmcalcs}. One can see from the Table that there is even disagreement on whether the 1-loop diagrams are finite or not, in that some feature a logarithmic divergence. These discrepancies (or those extant at the time) led Boudjema {\em et al.} \cite{boudjema} (who themselves did the computation with a number of different regulators and obtained a number of different results; we quote only the result obtained using dimensional regularization in the Table) to conclude that the EDM cannot be calculated in this way and depends intrinsically on the details of the UV physics. 
We instead take the view that, given the arguments in the Introduction, it must be possible to calculate in this way, and so we seek to resolve the discrepancies. To do so, we consider each calculation in order of increasing size.

Hoogeveen \cite{hoogeveen} gets zero, using a momentum cut-off regulator, but remarks himself that the result is dependent on the definition of the loop momentum that appears in the integral,\footnote{That is, $\int^\La \dfk \frac{1}{k^2 - m^2} \neq \int^\La \dfkp \frac{1}{(k^\prime + P)^2 - m^2}$, because the cut-off breaks the integral's invariance under shifts of $k$.} and is therefore suspect. Subsequently Boudjema {\em et al.} \cite{boudjema}  showed that Hoogeveen's result can be reconciled with others via suitable shifts of loop momentum. But as we shall see below, any result obtained with a momentum cut-off is unreliable.

The next smallest is the result of Novales-S\'anchez \& Toscano \cite{novalessanchez}, obtained using \msbar, which is suppressed by a factor $m_f^2/m_W^2$, where $m_f$ is a light quark mass, relative to other results. In \cite{novalessanchez}, two purported explanations are given for the suppression. The first is custodial symmetry. The operator $\op_W$ is indeed invariant under a $SU(2)_L \times SU(2)_R$ symmetry under which the $W$ boson transforms as a $(\mathbf{3},\mathbf{1})$, but this cannot explain the suppression. This is easily seen in the following way. In the limit $m_u = m_d$, custodial symmetry becomes an {\em exact} symmetry of all the interactions appearing in the diagrams of Fig.~\ref{fig:nedm}. If custodial symmetry suppresses the EDM, then the result quoted in \cite{novalessanchez} should vanish in this same limit, but it does not. In fact, custodial symmetry does not imply a constraint on either diagram; rather it relates the two diagrams, which sum to zero in the limit $m_u =m_d$. (To see this, consider the element of $SU(2)_L \times SU(2)_R$ given in the fundamental representation by $L=R= e^{ \frac{i \pi \sigma^1}{2} } = i \sigma^1$. Up to an overall phase, this effects the transformation $W^1 \rightarrow W^1, \, W^2 \rightarrow -W^2, \, W^3 \rightarrow -W^3, \, u_L \rightarrow d_L, \, d_L \rightarrow u_L$. The end result is that one of the charged current vertices picks up a minus sign when transforming from left to right in Fig.~\ref{fig:nedm}.) The second explanation invokes the decoupling theorem \cite{Appelquist:1974tg}, which, applied to the situation at hand, states that all effects of $W$ bosons on low energy physics (such as the neutron EDM) should decouple in the limit $m_W \rightarrow \infty$. Whilst this is quite true, one cannot take the limit $m_W \rightarrow \infty$ within an EFT without simultaneously taking $\Lambda \rightarrow \infty$. Thus, the $\Lambda^{-2}$ in the EDM result guarantees that the decoupling theorem is obeyed, without the need for an extra factor of $m_W^{-2}$. In fact, it turns out that the calculation in \cite{novalessanchez} is erroneous.

The remaining calculations give results that are at least consistent with a na\"{\i}ve estimate. Since this already implies a very strong bound on $\kg$, the reader's interest may well begin to wane at this point.
Nevertheless, it is perhaps worth persevering, in that there is a valuable EFT lesson to be learnt in resolving the remaining discrepancies of Table~\ref{tab:nedmcalcs}. 

The discrepancies are catalogued in exhausting detail in \cite{boudjema}, where we learn not only that a momentum cut-off regulator leads to an ambiguous result, but also that other regulators (namely dimensional regularization, a Pauli-Villars regulator, and a form factor) all lead to different results. We must conclude either, as in \cite{boudjema}, that the result is ambiguous
(in which case the edifice of QFT collapses) or that ({\em pace} Orwell), whereas in QFT all regulators are equal, in EFT some regulators are more equal than others.

In fact, the latter conclusion is the correct one.
The reason is that, in order for an EFT to be of any use, one must find a way in which to control the infinity of operators that appear, once one gives up the criterion of renomalizability. At tree-level this can be achieved by classifying the possible operators by their scaling dimensions (in energy) and truncating at a certain order. This guarantees that computations at energies below the cut-off will be accurate up to a given power of the energy divided by the cut-off. But once we include loops, we encounter the usual problem of UV divergences. In renormalizable QFT, we may regularize these however we like, for example via an explicit energy cut-off $\La$. But in EFT, such a cut-off undoes the tree-level momentum expansion: the infinity of higher dimension operators, though suppressed at low energies, are unsuppressed in loops, where energies up to the cut-off are allowed. 

This phenomenon is easily illustrated (see {\em e.g.} \cite{manohar}) by considering 1-loop corrections to the dimension-4 operator $\lambda \phi^4$ in the simplest example of a scalar field theory where some heavy particles of mass $\La$ have been integrated out. The EFT Lagrangian is
\begin{equation*}
\lag = -\frac{1}{2}\phi(\pt^2 + m^2)\phi - \frac{1}{4!} \lambda \phi^4 \\
- \frac{1}{6!} \frac{c_6}{\Lambda^2} \phi^6 - \frac{1}{2 \cdot 4!} \frac{c_8}{\Lambda^4} \phi^4 (\pt \phi)^2 - \ldots,
\end{equation*}
where the dimensionless coefficients $c_6, c_8, \ldots$ are $O(1)$. A \emph{mass dependent} regulator, such as a momentum cut-off at the limit of validity of the theory $\La$, gives loop diagrams of similar size from all operators.\footnote{An apparently simple solution to this problem would be to use a lower cut-off $\La^\prime < \La$ for the loop integral. But doing so generates operators with derivatives of size $\frac{\pt}{\La^\prime}$ under the renormalization group flow, thereby reducing the regime of validity of the EFT as a whole to $p \lesssim \La^\prime$.} Indeed,
\begin{align*}
\delta \lambda_{\text{1-loop}} &\supset \frac{c_6}{\La^2} \int^\La \frac{\dif^4 k}{(2\pi)^4} \frac{1}{k^2-m^2} \sim \frac{c_6}{\La^2} \frac{\La^2}{16 \pi^2} \sim \mathrm{O}(1), \\
\delta \lambda_{\text{1-loop}} &\supset \frac{c_8}{\La^4} \int^\La \frac{\dif^4 k}{(2\pi)^4} \frac{k^2}{k^2-m^2} \sim \frac{c_8}{\La^4} \frac{\La^4}{16 \pi^2} \sim \mathrm{O}(1), \; \mathit{\&c.}  
\end{align*}
Thus we find that predictivity is lost once again using such a cut-off, since we need to consider loops containing all operators to calculate at any given order in the momentum expansion of the Lagrangian.

The solution is a simple one: A \emph{mass independent} regulator, such as dimensional regularization with \msbar, gives no problematic powers of the renormalization scale in the numerator. Indeed, the only mass scales that can appear in the numerators of diagrams correspond to light masses or momenta, with the renormalization scale appearing only in logarithms. For the EFT of a scalar, for example,
\begin{align*}
\frac{c_6 \mu^{2\epsilon}}{\Lambda^2} &\int \frac{\dif^{4-\epsilon} k}{(2\pi)^{4-\epsilon}} \frac{1}{k^2-m^2} \sim \frac{c_6}{\Lambda^2} \frac{m^2}{16 \pi^2} \frac{1}{\epsilon} - \frac{c_6}{\Lambda^2} \frac{m^2}{16 \pi^2} \ln( \frac{m^2}{\mu^2} ), \\
\frac{c_8 \mu^{2\epsilon}}{\Lambda^4} &\int \frac{\dif^{4-\epsilon} k}{(2\pi)^{4-\epsilon}} \frac{k^2}{k^2-m^2} \sim \frac{c_8}{\Lambda^4} \frac{m^4}{16 \pi^2} \frac{1}{\epsilon} - \frac{c_8}{\Lambda^4} \frac{m^4}{16 \pi^2} \ln( \frac{m^2}{\mu^2} ), \; \mathit{\&c.} 
\end{align*}

A mass independent scheme thus preserves the original momentum expansion: contributions from higher dimension operators are suppressed, even in loops. If we consider all operators up to dimension $d$, we are guaranteed a result accurate to $\bo \left( (p/\La) ^{d-4} \right)$, where $p$ is the momentum scale of the process, at any loop order.

Thus, the results in Table~\ref{tab:nedmcalcs} that are regularized with a form factor or a cut-off are wrong, for they will receive corrections of comparable sizes from the $d > 6$ operators ignored in the EFT, and also from higher loop orders. Dimensional regularization is the way forward, and by doing it we independently obtain the same result as Ref.~\cite{boudjema} does in this case, namely
\begin{equation}
d_f = m_f \kg \frac{g_W^2}{64 \pi^2}.
\label{eqn:rightanswer}
\end{equation}

This may be translated into a bound on $\kg$. Expressing a fermion $f$'s EDM operator as $-\frac{1}{2}i d_f \overline{\psi}_f \sigma^{\mu\nu} \psi_f \tilde{F}_{\mu\nu}$, experiment gives $\abs{d_n} < 2.9 \times 10^{-26} e \, \mathrm{cm}$ at 90\% C.L. \cite{baker2006}. We use the form factors of \cite{dib2006} to convert this into quark EDM bounds: $d_n \sim 1.77d_d - 0.48 d_u$. The result (\ref{eqn:rightanswer}) gives $\abs{\kg} \lesssim 6 \times 10^{-8} \Gev^{-2}$.\footnote{For a discussion of the RG corrections to these bounds, see \cite{dekens13}.}

Evidently, barring implausible cancellations against contributions coming from other higher dimension operators, we cannot hope to see an effect from $\op_\gamma$ in searches at the LHC. Nonetheless, authors have advocated searching for its sister $\op_Z$ as a potential source of $CP$ violation. As we have already described, to have a small $\op_\gamma$ and sizeable $\op_Z$ is to introduce explicit \ewgg violation of the form $\op_B$, since
\begin{equation*}
\kg \op_\gamma + \al_Z \op_Z = (c_W \al_Z + s_W \kg) \op_{W} + (c_W \kg - s_W \al_Z) \op_B .
\end{equation*} 
As we show in the next Section, doing so inevitably lowers the EFT cut-off below $\La_B$, where $\kb \equiv \La_B^{-2}$.

\section{Cut-offs in an EFT with $\op_B$}
\label{sec:unitcutoff}
Having discounted $\op_W$ as a means of getting a visible $WW\tilde{Z}$ effect at the LHC, we now consider obtaining it from the operator $\op_B = {W^+}^\mu_\nu {W^-}^\nu_\lambda \tilde{B}^{\lambda}_\mu$. This latter operator does not respect the \ewgg gauge symmetry of the renormalizable SM. Now, this invariance is clearly not a {\em sine qua non} --- it is violated, for example, in the mass terms of the $W$ and $Z$ bosons in a Higgsless model. However, then, as now, we must ask what the cut-off of the theory is. More precisely, beyond what energies does the theory lose predictivity or consistency?

This is not necessarily a straightforward question to answer. Indeed, the introduction of just one \ewggnospace-violating term engenders the appearance of all others via loop effects, and any one of these could jeopardise consistency and predictivity. Moreover, to compute the cut-off one must, in principle, analyze all possible scattering amplitudes in the theory, comparing contributions at arbitrary orders in the perturbation expansion.\footnote{We remind the reader that there are really two expansions: one in energies and one in loops. However, the two expansions are related by the fact that the loops lead to integrals that themselves require an energy cut-off.}

\begin{figure}
\centering
\vspace{5mm}

\begin{tabular}{c c c}

\begin{fmffile}{smmixingtree_feynmp}
\begin{fmfgraph*}(30,30)
\fmfleft{b}
\fmfright{wp,wm}
\fmf{photon,tension=3}{b,dummy1,mix,dummy2,v1}
\fmf{photon}{wp,v1,wm}
\fmfv{decor.shape=cross}{mix}
\fmflabel{$B$}{b}
\fmflabel{$W^+$}{wp}
\fmflabel{$W^-$}{wm}
\fmfv{label=$W^3$,label.angle=90}{dummy2}
\end{fmfgraph*}
\end{fmffile}

&
\hspace{15mm}
&

\begin{fmffile}{blobtree_feynmp}
\begin{fmfgraph*}(30,30)
\fmfleft{b}
\fmfright{wp,wm}
\fmf{photon}{b,v1}
\fmf{photon}{wp,v1,wm}
\fmflabel{$B$}{b}
\fmflabel{$W^+$}{wp}
\fmflabel{$W^-$}{wm}
\fmfblob{20}{v1}
\end{fmfgraph*}
\end{fmffile}

\end{tabular}
\vspace{5mm}
\caption{\label{fig:wwb} Contributions to $W^+ W^- B$ scattering in the SM with the operator $\op_B = {W^+}^\mu_\nu {W^-}^\nu_\lambda \tilde{B}^{\lambda}_\mu$ (denoted by a shaded blob) added. The `$\times$' denotes the mixing between $W^3$ and $B$ in the SM.}
\end{figure}
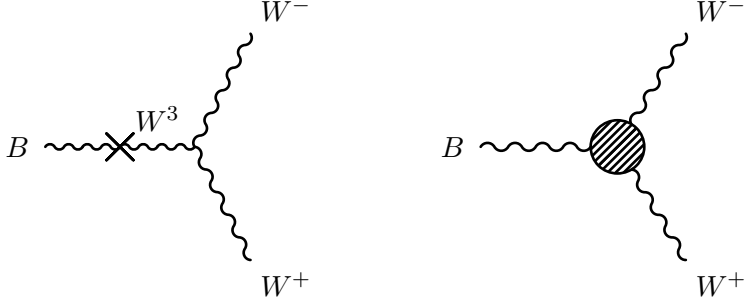

In fact, in the case at hand, we can see that the cut-off must be lowered by means of a trivial calculation. Indeed, consider the diagrams of Fig.~\ref{fig:wwb}, which represent contributions to $W^+ W^- B$ scattering at different orders in the momentum expansion. For simplicity, we assume that the gauge couplings, and hence the masses and mixings are small: this suffices to derive the functional dependence of the cut-off.
The first diagram, which is at leading order in the momentum expansion, arises in the SM from the 3-point Yang-Mills vertex with an insertion of the $W^3$-$B$ mixing operator $\frac{1}{4} g g^\prime v^2 W^3_\mu B^\mu$.
It has size $g g^\prime v^2 \cdot \frac{1}{p^2} \cdot g p$,
whereas the second diagram involves $\op_B$, arises at the next order in the momentum expansion, and has value
$ \kb p^3 $.
The momentum expansion breaks down roughly at a scale $\La$ where these terms become of equal size, namely when 
\begin{equation}
\La \sim \left( g^2 g^\prime v^2\La_B^2\right)^\frac{1}{4}.
\end{equation}
It is thus abundantly clear that the cut-off that follows from the presence of $\op_B$ is not the na\"{\i}ve $\La_B$, but is suppressed. The suppression comes not just from the ratio $v/\La_B$, but also from factors of the gauge couplings.
Thus, for a given size of $\kb$ (which sets the size of new physics effects of $\op_B$ at the LHC and elsewhere), we find a cut-off that is far lower than the na\"{\i}ve one. 

Derived in this way, our result has the air of prestidigitation and it would be desirable to obtain it in a more perspicuous fashion. This too is easily done, by using a formalism that is completely equivalent, but in which \ewgg is manifest, albeit non-linearly realised \cite{Coleman:1969sm,ccwz69}. To do so, we define the sigma model field
$\Sigma = e^\frac{i \pi^a \sigma^a}{2 v}$, where $\pi^a$ are three Goldstone boson fields and $\sigma^a$ are the Pauli matrices. Under an \ewgg transformation, $\Sigma \rightarrow U_L \Sigma U_Y^\dagger \equiv e^{i \al_L^a \frac{\sigma^a}{2}} \Sigma e^{- i \al_Y \frac{\sigma^3}{2}}$. Our original EFT, {\em viz.}
\begin{equation}
\lag = - \frac{1}{4} W^{a \mu\nu} W^a_{\mu\nu} - \frac{1}{4} B^{\mu\nu} B_{\mu\nu} - m_W^2 W^{+\mu} W^-_\mu - \frac{1}{2} m_Z^2 Z^{\mu} Z_\mu - i\frac{1}{\La_B^2} {W^+}^\mu_\nu {W^-}^\nu_\lambda \tilde{B}^\lambda_\mu  + \dots
\label{eqn:ungauged}
\end{equation}
can then be re-written as
\begin{equation}
\lag = - \frac{1}{4} W^{a \mu\nu} W^a_{\mu\nu} - \frac{1}{4} B^{\mu\nu} B_{\mu\nu} + \frac{v^2}{2} \Tr ((D_\mu \Sigma)^{\dagger} (D^\mu \Sigma) ) - \frac{1}{4\La_B^2} \Tr ( \Sigma^\dagger {W}^\mu_\nu {W}^\nu_\lambda \Sigma \tilde{B}^\lambda_\mu ) + \dots,
\label{eqn:gauged}
\end{equation}
where $D_\mu \equiv \partial_\mu + igW_\mu^a \frac{\sigma^a}{2} + ig^\prime B_\mu \frac{\sigma^3}{2}$. To see that the two EFTs are equivalent, it suffices to fix the gauge $\Sigma = 1$ in (\ref{eqn:gauged}). 

As has been emphasized, {\em e.g.} in \cite{arkanihamed02}, the formulation (\ref{eqn:gauged}) is far more convenient for the purposes of extracting the EFT cut-off $\La$. Indeed, the cut-off is finite because scattering amplitudes involving longitudinal gauge boson polarizations grow with the energy. But at high energies we may invoke the 
Goldstone boson equivalence theorem and replace the longitudinal gauge bosons by Goldstone bosons. Thus, the cut-off can be extracted by analyzing the Goldstone bosons alone. Now, we have known for a long time how to extract the cut-off of such a non-linear sigma model. The result is that, if we write the EFT as
\begin{equation}\label{eq:pc}
\mathcal{L} = \La^2 v^2 F \left( \frac{\partial}{\La}, \frac{gA}{\La}, \Sigma \right)
\end{equation}
(where we have generically indicated a gauge field and its coupling by $A$ and $g$, respectively), then the theory has cut-off $\Lambda$, where $\Lambda \lesssim 4 \pi v$ \cite{Manohar:1983md,Georgi:1986kr}.\footnote{Note that the power counting given in (\ref{eq:pc}) only applies for terms where the symmetry is non-linearly realized. It does not apply to the gauge kinetic terms or $\op_W$, for example.} This result immediately tells us that the coefficient $\kb$ in
(\ref{eqn:gauged}) is given by 
 \begin{equation} \label{eqn:truesize}
\kb \sim \frac{g^2 g^\prime v^2}{\La^4},
\end{equation}
where $\La$ is the true cut-off of the theory.\footnote{Ref. \cite{lep96} obtains a similar result for both $\kg$ and $\kz$ ``within the nonlinear realization scenario''; in fact, it is only correct for $\kb$. Moreover, our discussion shows that there is no alternative scenario.} Once again, we see that cut-off is not $\La_B \equiv 1/\sqrt{\kb}$.

Thus far, we have avoided referring to the Higgs doublet $H$, but it is straightforward to include it in the discussion. If the Higgs is present (and the LHC suggests that it is) then we have one more field that can be included in our EFT.\footnote{Though, as we shall see, the cut-off needed to generate observable effects from $\op_B$ at the LHC is so low that we are almost justified in using a description in which a 125 GeV Higgs is integrated out!} This can be done straightforwardly by the replacement $\Sigma \rightarrow H$ in (\ref{eqn:gauged}). The Higgs field unitarizes gauge boson scattering and so the cut-off of the resulting EFT can be made arbitrarily large. Nevertheless, the operator $\kb^\prime \Tr ( H^\dagger {W}^\mu_\nu {W}^\nu_\lambda H \tilde{B}^\lambda_\mu )$ is of dimension eight, and the resulting $WWZ$ operator has coefficient $\sim \kb^\prime v^2$. Yet again, observability of the effects of $\kb^\prime \neq 0$ implies new physics at a low energy scale.

Now that we have some confidence in our result, we should ask just how low the cut-off must be, in order for us to have a chance of seeing the effects of $\op_Z$ at the LHC.
To date, there have been no dedicated ATLAS or CMS searches for such operators,\footnote{Although DELPHI has a bound of $\kb m_W^2 = -0.08 \pm 0.07$ \cite{delphi08}.} and so we content ourselves with re-intepreting the projections of  
\cite{han} for searches for $CP$ violation via the operator $\kz \op_{Z}$, in the light of our results. A $CP$-odd observable is constructed using the momenta of the leptonic decay products of a $W^+W^-$ pair. Using reasonable cuts the authors find, for the SM plus $\op_{Z}$, with $100 \, \mathrm{fb^{-1}}$ of data, the 14 TeV LHC is sensitive at the $7 \sigma$ level to $ \abs{ \kz m_W^2} = 0.1$. The non-zero contribution to the $CP$-odd observable will come from the interference between SM and $\op_Z$ amplitudes, a term linear in $\kz$, whereas the statistical fluctuations in the number of events (i.e. the size of a $\sigma$) come predominantly from the constant SM cross-section. Hence for a $5\sigma$ detection we require $\abs{ \kz m_W^2} \gtrsim  \frac{5}{7} \times 0.1$.

Given our size estimate (\ref{eqn:truesize}) for $\kb$ in terms of the true cut-off, we conclude that this maximum of sensitivity corresponds to a theory with cut-off $\La \sim 170 \Gev$. An electroweak sector that becomes strongly coupled at this energy would contain an infinite set of non-SM effective operators, each with $\bo(1)$ effects on scattering amplitudes at momenta $\sim 170 \Gev$, and by extension $\bo(1)$ contributions to electroweak precision tests. Needless to say, such large effects are absent in existing measurements. We infer from this absence that the effects of $\op_Z$ are unlikely to be seen at the LHC.

\section{Conclusions}
The baryon asymmetry of the Universe requires new, $CP$-violating physics, which we would dearly like to discover (or at least constrain) at the LHC.  Here we have analysed the prospects for operators involving three electroweak gauge bosons in an effective field theory approach. At dimension six, there is just one such operator, $\op_W$, which generates corrections at 1-loop to the EDM of the neutron. We have resolved the discrepancies in previous attempts to compute this correction. We find a sizeable contribution and thus we conclude that its coefficient must be negligibly small, as far as LHC physics is concerned.

This conclusion could be evaded by imagining that there are contributions to the neutron EDM from other higher dimension operators in the EFT. Whilst this is certainly possible, it would require a fantastic tuning between the coefficients of such operators. It is important to stress that such a  tuning would be just as miraculous, qualitatively if not quantitatively, as that which is needed to keep the Higgs mass parameter small in the SM equipped with a large cut-off: somehow, the parameters of the UV physics must be delicately chosen so as to effect a cancellation in a low energy property of the neutron. Why on Earth would the theory arrange itself `just so' as to hide itself in the experiments that we happen to be currently able to do? 

We have also shown that another possible loophole, namely that operators involving $WWZ$ and $WW\gamma$ can be generated independently, is bogus. Any such split is due to an operator that is really of dimension eight and so the cut-off corresponding to new physics effects of a given size is much lower than that which is na\"{\i}vely assumed. We estimate that with $100 \, \mathrm{fb}^{-1}$ of data at a 14 TeV LHC, we cannot probe much beyond a cut-off of $170 \Gev$. If strongly coupled physics did appear in the gauge boson sector at such a scale, we would surely have already seen it elsewhere. Thus, there are more promising avenues of discovery at the LHC.

It is possible that experimentalists will not heed our advice, and will stubbornly pursue these searches. (Perhaps it is for the best that they do, given theorists' recent track record of predictions.) If they do,
it is important that the results are interpreted with care. A particular danger in testing EFTs at high energy colliders is that some events may be outside the regime of validity of the theory, {\em viz.} at energy scales beyond the cut-off. Such a theory can neither be excluded nor `included' by an experimental search using those events {\em a priori}, given that it does not make a prediction in that regime. At least for the purposes of exclusion, a meaningful (and conservative) strategy would be to assume that the new physics contribution to the cross-section becomes negligible above the cut-off. Unfortunately, in current searches for $CP$-even operators involving triple gauge bosons, neither ATLAS nor CMS follow this strategy. CMS \cite{cmswgamma,cmszz} does not take the cut-off into account at all, whereas  ATLAS uses a form factor $\frac{1}{ (1 + \frac{\hat{s}}{\Lambda^2} )^n }$ with positive integer $n$ and a borderline value of $\La = 2,3,6 \Tev$ \cite{atlaszz,atlaswgamma,atlasww}. $\La$ is chosen such that the theory is always unitary for all the `included' values of coefficients, whereas for most of the excluded values the theory is not unitary at sufficiently high scattering momenta. This gives the unusual scenario where the effect of the `included' parameter points is potentially underestimated (since the cross-section above the cut-off falls sharply to zero), and the excluded points' effects are overestimated --- the very opposite of a conservative search. We recommend that, in future searches, the cut-off be allowed to float with the parameters, in such a way that unitarity is guaranteed at all parameter space points.

\section*{Acknowledgements}
BG acknowledges
the support of the Science and Technology Facilities Council, the
Institute for Particle Physics Phenomenology, and King's College,
Cambridge and thanks CERN, the Galileo Galilei Institute for Theoretical Physics, and Nordita for hospitality. DS acknowledges the support of the Science and Technology Facilities Council, as well as Emmanuel College, Cambridge. We are grateful to H.~Novales-S\'{a}nchez for verifying our calculation of the loop diagrams in \S\ref{sec:nedmcalc}.

\bibliographystyle{JHEP}
\bibliography{eft_gauge_unitarity_cutoff}

\end{document}